\documentclass[graybox]{svmult}
\usepackage{newtxtext,newtxmath}
\usepackage{helvet}         
\usepackage{courier}        
\usepackage{type1cm}        
\usepackage{makeidx}         
\usepackage{graphicx}        
\usepackage{multicol}        
\usepackage[bottom]{footmisc}
\usepackage[T1]{fontenc}
\makeatletter
\renewcommand{\chap@hangfrom}[1]{}
\makeatother
\makeindex             

\begin{document}

\title{Probing Atmospheric Escape Through the Near-Infrared Helium Triplet\thanks{Parts of this chapter have been adapted from the PhD thesis: The Design and Development of NIGHT: a High-Resolution, Near-Infrared Spectrograph for Exoplanetary Atmospheric Escape Studies via Helium Triplet Spectroscopy}}
\author{Farret Jentink, C.\orcidID{0000-0001-9688-9294},\\ Bourrier, V.\orcidID{0000-0002-9148-034X} and\\ Carteret, Y.\orcidID{0000-0002-6159-6528}}
\institute{Farret Jentink, C. \at Université de Genève, Département d'Astronomie, \email{casper.farret@unige.ch}
\and Bourrier, V. \at Université de Genève, Département d'Astronomie \and Carteret, Y. \at Université de Genève, Département d'Astronomie}

\setcounter{chapter}{0}
\maketitle
\vspace{-0.5cm}

\abstract{
The most productive tracer of exoplanetary atmospheric escape is the measurement of excess absorption in the near-infrared metastable helium triplet during transits. Atmospheric escape of a close-in planet’s atmosphere plays a role in its evolutionary pathway, but to which extent remains unknown. It could possibly explain demographic features like the radius valley and Neptunian desert. We will describe the development of instrumental, reduction, and modelling techniques to study exoplanetary atmospheric escape, focusing on the helium triplet. One such developments is the NIGHT spectrograph, intended to provide the first survey of escaping atmospheres. NIGHT spectra will be processed with ANTARESS, a state-of-the art workflow to reduce high-resolution spectral time-series of exoplanet transits, and compute transmission spectra in a robust and reproducible way. Transmission spectra contain the potential signature of the planetary atmosphere as well as distortions induced by the occultation of local regions of the stellar surface along the transit chord. 
Transmission spectra cannot be corrected for those stellar distortions without biasing the planetary signal, and must instead be directly interpreted using a numerical model like the EvE code, which generates realistic stellar spectra that account for the system's 3D architecture, the planet's atmospheric structure, and its local occultation of the stellar disc. This global approach, from the measurement and computation of transmission spectra to their interpretation, will be a legacy of \textit{PlanetS}, becoming the standard procedure to study high-resolution spectroscopy of planetary transits.}

\section{Introduction}

Observations of exoplanetary atmospheres offer a unique insight into the physical and chemical properties of these distant worlds, enabling a detailed characterization of their composition, atmospheric dynamics, and internal structure.

\subsection{Planetary Atmospheric Escape}

The detection of 51 Pegasi b in 1995 \cite{mayor1995} created an entirely new subfield of astronomy, ushering in the era of exoplanet science. Before this breakthrough, our understanding of planetary systems was limited to the Solar System. While it had already revealed the existence of planets spanning a large range of masses and radii, it had not prepared us for the existence of massive gas giants on very short orbits.  

The field of exoplanet detection initially relied on high-resolution spectroscopy to measure stellar radial velocities, where periodic variations constrained the period and mass of orbiting planets. To constrain the bulk planetary density, only the radius of the planet remained to be measured. A significant advancement came in 1999 with the first observed exoplanetary transit of HD209458 b \cite{charbonneau2000}, which opened the avenue for the characterization of exoplanets. This discovery was followed by another breakthrough in 2002 and 2003, when transmission spectroscopy of the same planet enabled atmospheric analysis through the identification of specific elemental absorption features \cite{charbonneau2002, Vidal2003}. \textit{Hubble Space Telescope/Space Telescope Imaging Spectrograph} (HST/STIS) observations revealed an excess absorption of 15\% in the ultraviolet (UV) Lyman-$\alpha$ line from a comet-like tail of hydrogen gas, providing the first proof of atmospheric escape from an exoplanet \cite{Vidal2003}.

\begin{figure}[!ht]
\includegraphics[width=1\textwidth]{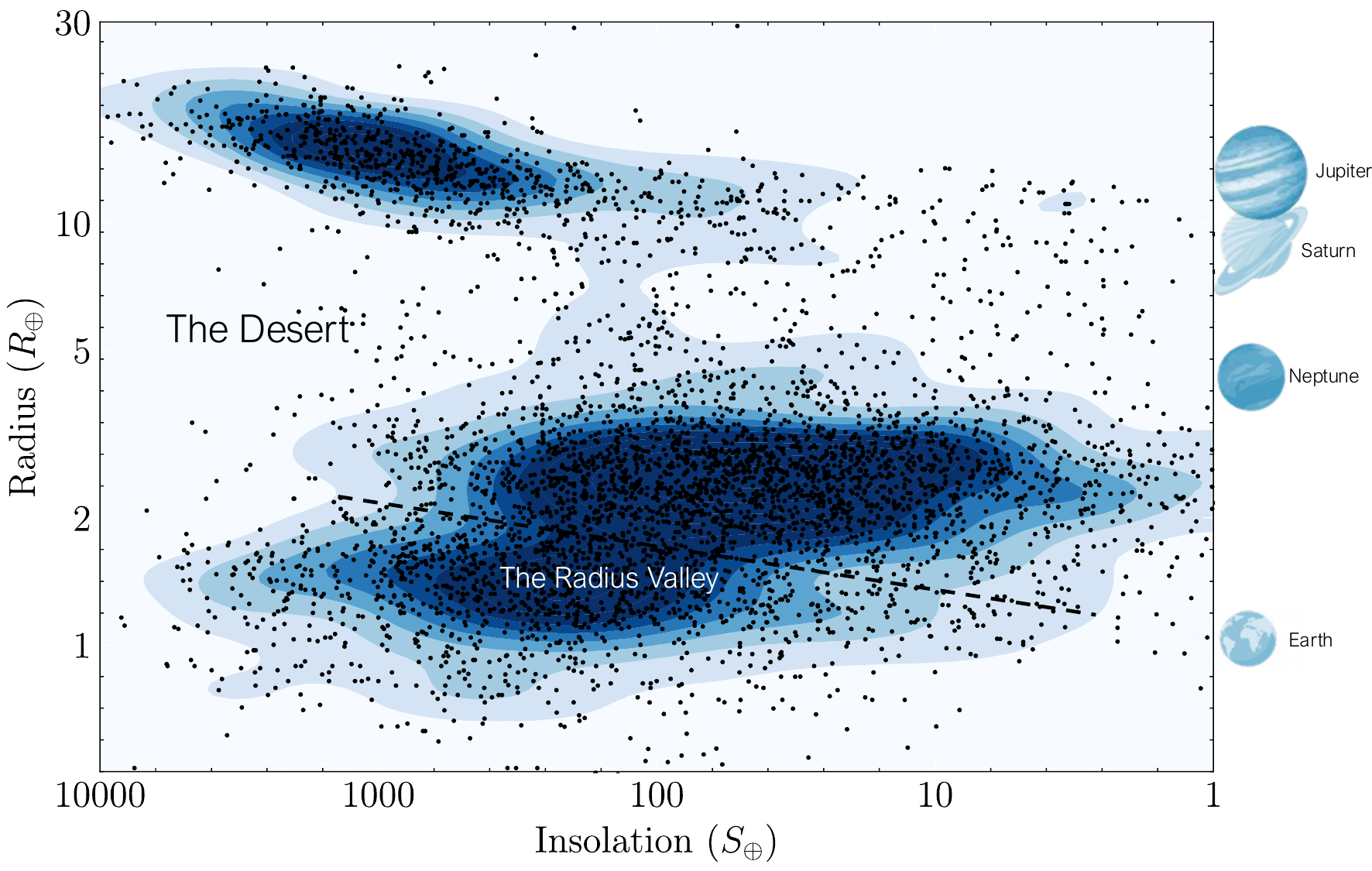}
\caption{The distribution of known transiting exoplanets on an insolation versus radius diagram. The contour highlights two notable regions of lower exoplanet density—the Neptunian Desert and the Radius Valley—whose formation mechanisms remain subjects of ongoing scientific debate. Notably, there is also the observable under-density of Neptune-sized planets on longer period orbits, also dubbed the Savannah. Both the Radius Valley and Savannah will require helium observations of longer-period and smaller-mass planets. Note that for lower insolations, the apparent under-density is highly likely a result of observational bias, as with the most productive detection methods it is inherently more difficult to detect long-period planets. For scale reference, four planets from our Solar System are shown on the right. Data taken from the NASA Exoplanet Archive and The Extrasolar Planets Encyclopaedia. \cite{akeson2013, schneider1996}}
\label{fig:insolation_distribution}
\end{figure}

Further observations with HST/STIS revealed atmospheric escape to be a widespread phenomenon among close-in exoplanets (e.g. \cite{lammer2003,lecav2010}). Important to note here is that the radial velocity and transit methods inherently favor the discovery of Hot Jupiters, massive planets orbiting extremely close to their host stars. The absence of smaller, longer-period planets in the early days was mainly a result of sensitivity and precision limitations, and poor temporal sampling. The intense stellar radiation that close-in gaseous planets receive makes them particularly susceptible to atmospheric loss. The development of more sensitive photometric and radial velocity instruments, and long-time baseline observations, has expanded our dataset to include rocky worlds and longer-period planets, for which it remains unclear whether atmospheric escape occurs. Nevertheless, compelling evidence suggests that atmospheric escape could play an important role in shaping the evolution of close-in exoplanets. An imprint of these evolutionary paths, although not yet fully understood, can be found in the insolation-radius distribution of all known transiting exoplanets, as seen in Figure~\ref{fig:insolation_distribution}. Despite the advances that were made over the past decades, current observational data remains insufficient to fully understand how atmospheric escape influences the evolutionary pathways of both gaseous and rocky planets. 

The present understanding of atmospheric escape in close-in gas-rich exoplanets suggests:
\begin{itemize}
    \item Hydrodynamic atmospheric escape is largely driven by X-ray and extreme-ultraviolet (XUV) radiation from host stars, heating up the atmosphere.\cite{owen2019,oklopcic2019}  However, it could also be partially driven by residual heat from the planet's formation for small sub-Neptune planets.\cite{gupta2019}
    \item Atmospheric escape happens for a large range of planetary types, at least from sub-Neptunes to Jupiters. For small enough planets, the lost mass can be enough to remove a significant fraction of their atmosphere over the course of millions of years. \cite{owen2018,bourrier2018}
    \item As such, the induced mass loss is expected to play a role \cite{lecav2004,bourrier2018,attia2021,koskinen2022} in the formation of the Neptunian desert (an apparent lack of Neptune-sized planets at high insolation, e.g., \cite{Lecav2007,Davis2009,Mazeh2016}) and possibly of the Neptunian savannah (a milder deficit of Neptune-sized planets at lower insolation, \cite{Bourrier2023}), and the radius valley, a statistically significant dip in the size distribution of known exoplanets at approximately 1.5-2 Earth radii \cite{fulton2017}, potentially indicating a transition zone between rocky super-Earths and gas-rich sub-Neptunes.\cite{ginzburg2018,venturini2020}
\end{itemize}

\subsection{The Helium Triplet as an Escape Tracer}
The first proposition of using spectroscopy during transits to determine the composition of exoplanet atmospheres was made by Saeger \& Sasselov \cite{seager2000}. Their work laid the foundation of the method used by Charbonneau et al. \cite{charbonneau2002} to detect the first exoplanet atmosphere, and is still used frequently as of this day. Notably, Seager \& Sasselov's work also highlighted the potential of a triplet of helium absorption lines in the near infrared ($\sim$10 833 A). This spectral feature arises from the metastable $2^3S_1$ state. With the state's exceptionally long decay time of $\tau \approx 7800$ s \cite{drake1971}, it was identified as a promising diagnostic tool for atmospheric escape characterization \cite{oklopcic2018}. Following the detection of HD209458 b's escaping atmosphere in Lyman-$\alpha$, Moutou et al. \cite{moutou2003} attempted but failed to detect the helium triplet for the same planet. It would take another 15 years before Spake et al. \cite{spake2018} successfully detected the helium triplet for the first time during an exoplanet transit of WASP-107b using HST. This breakthrough was quickly followed up with a ground-based confirmation by Allart et al. \cite{allart2019}. Both teams included contributions from \textit{PlanetS}.

Despite not probing the same layers of an exoplanet atmosphere, both neutral hydrogen through Lyman-$\alpha$ and helium through the triplet have proven to be effective tracers of atmospheric escape. The helium triplet lines, being situated in the near-infrared, do not suffer from absorption from the interstellar medium and Earth's atmosphere, and benefit from a brighter stellar continuum, thus offering greater sensitivity. This also reflects in the number of Lyman-$\alpha$ transit detections published to this date, which remain limited to just a handful. Unlike neutral hydrogen, the density of metastable helium is extremely low, typically only a few hundred particles per cm$^3$ at most because only a limited number of processes can excite helium from its ground state to the metastable state. Thus, it does not provide a direct measurement of the atmospheric mass-loss rate, and interpretations must rely on models that incorporate a self-consistent treatment of upper-atmospheric chemistry. As mentioned previously, metastable helium naturally decays to the ground state, limiting the spatial extent of the metastable helium exosphere compared to that of neutral hydrogen. Consequently, the region where most metastable helium absorption is located remains mostly collisional because of the high density of the planetary outflow \cite{Wang2021}. Very extended outflows, such as described in \cite{Zhang2023,Gully_Santiago2024, allart2025_2} are produced by massive roche-lobe overflows around large stars. In this regime, the atmospheric escape remains collisional on very large scales, this is supported by hydrodynamical simulations (e.g., \cite{MacLeod2022,macleod2024streams}). However, for smaller mass-loss rates, the collision between particles can be sparse which forms an exosphere best described by a particle based model. Metastable helium traces dense regions within a collisional regime whereas it is less clear for neutral hydrogen observed through Lyman-$\alpha$.

\section{The NIGHT Project: A Dedicated Instrument and Survey to study Atmospheric Escape}
\subsection{Scientific Rationale for NIGHT}
\begin{figure}[!ht]
\centering
\includegraphics[scale=.6]{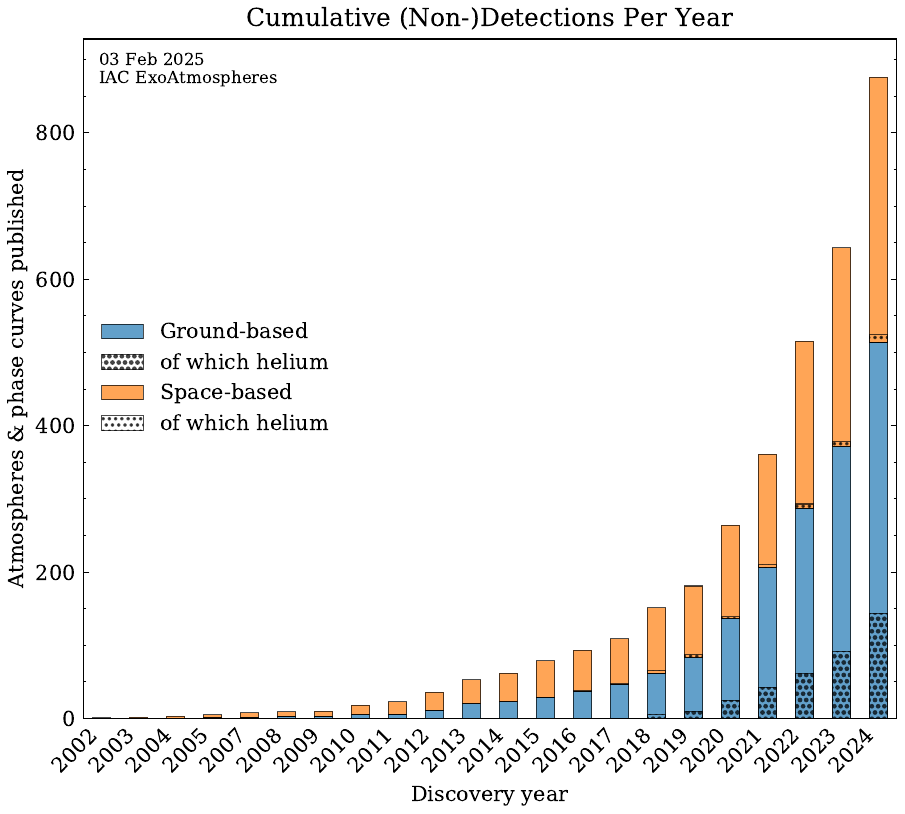}
\caption{The number of atmospheric detections of exoplanets published over time as extracted from the IAC ExoAtmospheres archive. The data includes (non-)detections and upper limits as all are relevant for constraining evolutionary pathways and interior composition models. Observations are split between ground-based and space-based, and helium detections are shown as subsets of these. It is important to note that this plot represents individual publications rather than unique detections. Multiple publications may report the same molecular species when detected by different instruments or observational campaigns, leading to apparent redundancy in the literature record. The ExoAtmospheres archive does not contain all available publications. Additionally, we would like to note that detections are more likely to be published than non-detections.}
\label{fig:atmosphere_detections_over_time}
\end{figure}

At the moment of writing, at least 28 detections, 10 non-detections, and 52 upper limits of helium in a large variety of exoplanets, ranging from sub-Neptunes to Jupiter-sized planets \cite{krishna2024,guilluy2023,allart2023,linssen2024}, have joined WASP-107b\footnote{From the IAC ExoAtmospheres archive}. \textit{PlanetS}-supported researchers have largely contributed to these studies, whose number is expected to continue to grow as more planets amenable to atmospheric characterization are detected. Figure~\ref{fig:atmosphere_detections_over_time} shows the rapid increase in published exoplanet atmospheres and phase curves, with a significant fraction of these being helium observations. 

Despite the rapid increase in the number of helium observations of exoplanet atmospheres, our current dataset remains insufficient to fully understand the role of atmospheric escape in planetary evolution. Significantly more observations are required to study the relations between escape and the properties of the planets and the stars hosting them. Although studies have identified correlations between the presence of escape and XUV flux \cite{krishna2024} and stellar mass \cite{guilluy2023,krishna2024}, the modelled mass loss rates show less clear trends \cite{guilluy2023}. The relationship between the rate of atmospheric escape and the characteristics of the planet's system probably involves a complex interaction among multiple parameters. A substantial increase in unbiased observational data is necessary to address these questions. Consequently, the conception of a specialized instrument and a survey for helium detection, termed NIGHT (\textit{The Near-Infrared Gatherer of Helium Transits}, \cite{farretjentink2024a,farretjentink2024b}) was envisioned early-on by \textit{PlanetS} researchers.

\subsection{Trade-Offs and Requirements}
Given NIGHT's highly specific science case, the instrument does not need to meet the broad set of science requirements typically expected for high-resolution spectrographs with a wide range of science cases. The helium triplet only covers a narrow wavelength band which--accounting for the planet velocities and some margin--translates to a 4nm bandpass over which NIGHT operates. Most science and instrument requirements were summarized in \cite{farretjentink2024a}. For completeness of this chapter we remind the key demands:

\begin{svgraybox}
    \begin{enumerate}
        \item Achieving a resolving power of R = 70,000 across the 1081-1085 nm wavelength range.
        \item Keep uncalibrated radial velocity (RV) drifts below the level of 40 m/s per night to ensure proper cross-calibration over the course of a transit. 
        \item Reach a sensitivity which allows for the detection of the helium triplet for at least 100 unique planet transits per year from a 2-meter class telescope.
        \item Although, technically not a requirement, we do emphasize that keeping NIGHT cost-efficient, simple, and compact is beneficial for the type of science that will be conducted. 
    \end{enumerate}
\end{svgraybox}

Throughout the last four years of the NCCR \textit{PlanetS}, with support from their Bonus Programme, the \textit{Trottier Institute for Research on Exoplanets} (iREx), the \textit{University of Geneva} and the \textit{Swiss National Science Foundation}, we developed NIGHT. At the moment of writing, the instrument is fast-approaching commissioning at the 152cm telescope at the Observatory of Haute Provence (OHP152) in France. Here, we detail the project's evolution, major design choices, and expected scientific returns from its first year at OHP152.

\subsection{A NIGHT Systems Overview}
\begin{figure}[!ht]
\includegraphics[width=0.9\textwidth]{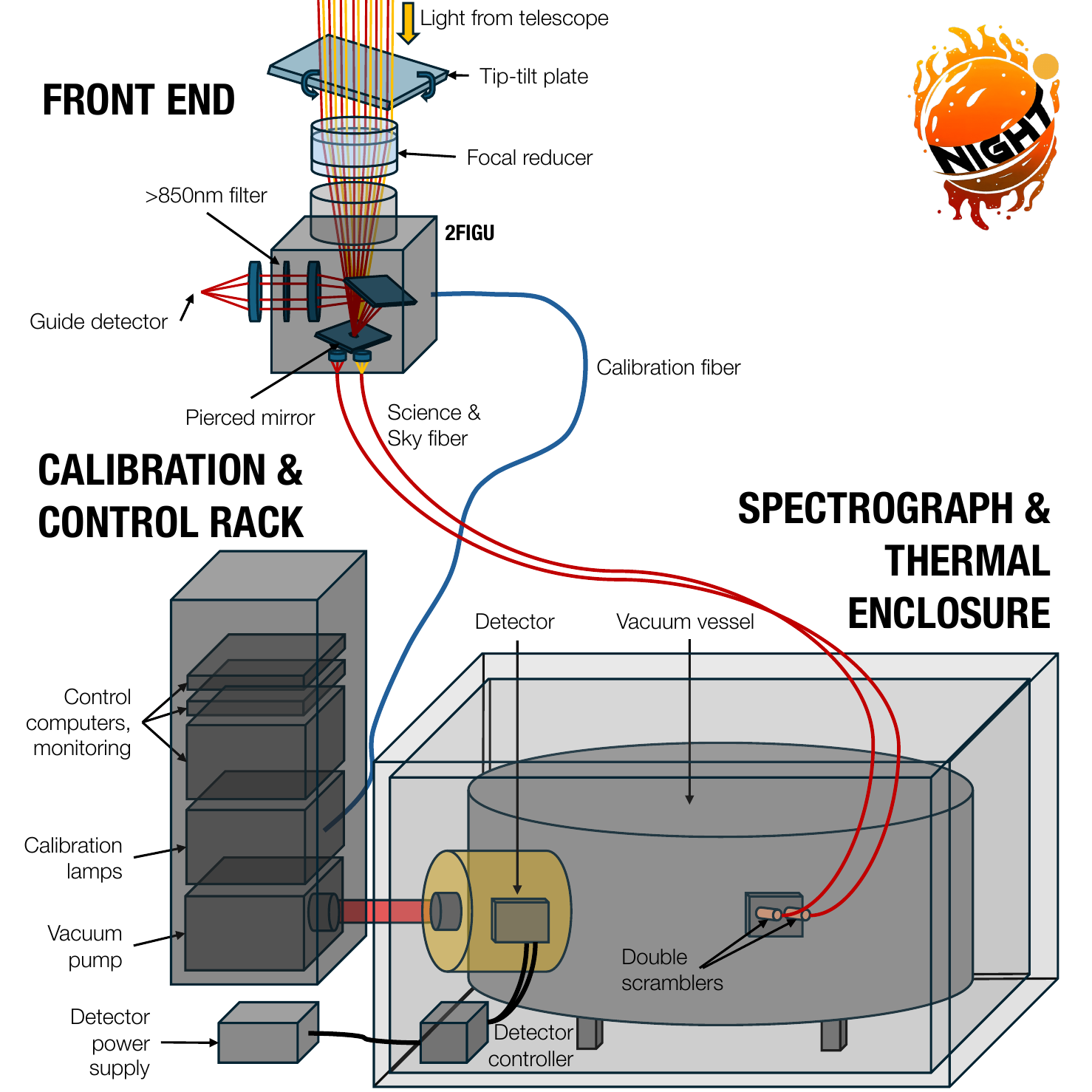}
\caption{This diagram illustrates the main subsystems comprising the NIGHT instrument. It consists of three main components: the front end unit, the calibration \& control systems, and the thermally-stabilized spectrograph. The front end interface, mounted to the telescope, couples stellar and sky light into two optical science fibers. This unit incorporates a tip-tilt correction capability for auto-guiding purposes and enables calibration light injection into either science fiber. The two science fibers extend from the front end to double scramblers positioned at the vacuum vessel entrance. The calibration and control cabinet encompasses the instrument’s control architecture -- including the central computer systems, thermal regulation for both vacuum vessel and detector, calibration sources, and the vacuum pump for the detector cryostat. This representation has been simplified for clarity; specific interfaces between components are excluded and the diagram is not to scale.}
\label{fig:systems_sketch_NIGHT}
\end{figure}

NIGHT consists of four main subsystems: the spectrograph with its thermal enclosure, the optical fiber train, the front end fiber injection and autoguide system, and the calibration and control cabinet. A simplified diagram showing most of the subsystems can be found in Figure~\ref{fig:systems_sketch_NIGHT}. Details have been omitted as they are beyond the scope of this chapter. A more detailed technical description of NIGHT's subsystems can be found in \cite{farretjentink2024b}. In the following subchapters we will provide a brief explanation of the various components of NIGHT, highlighting the most important design choices.

\subparagraph{The Spectrograph}

\begin{figure}[!ht]
\centering
\includegraphics[width=0.65\textwidth]{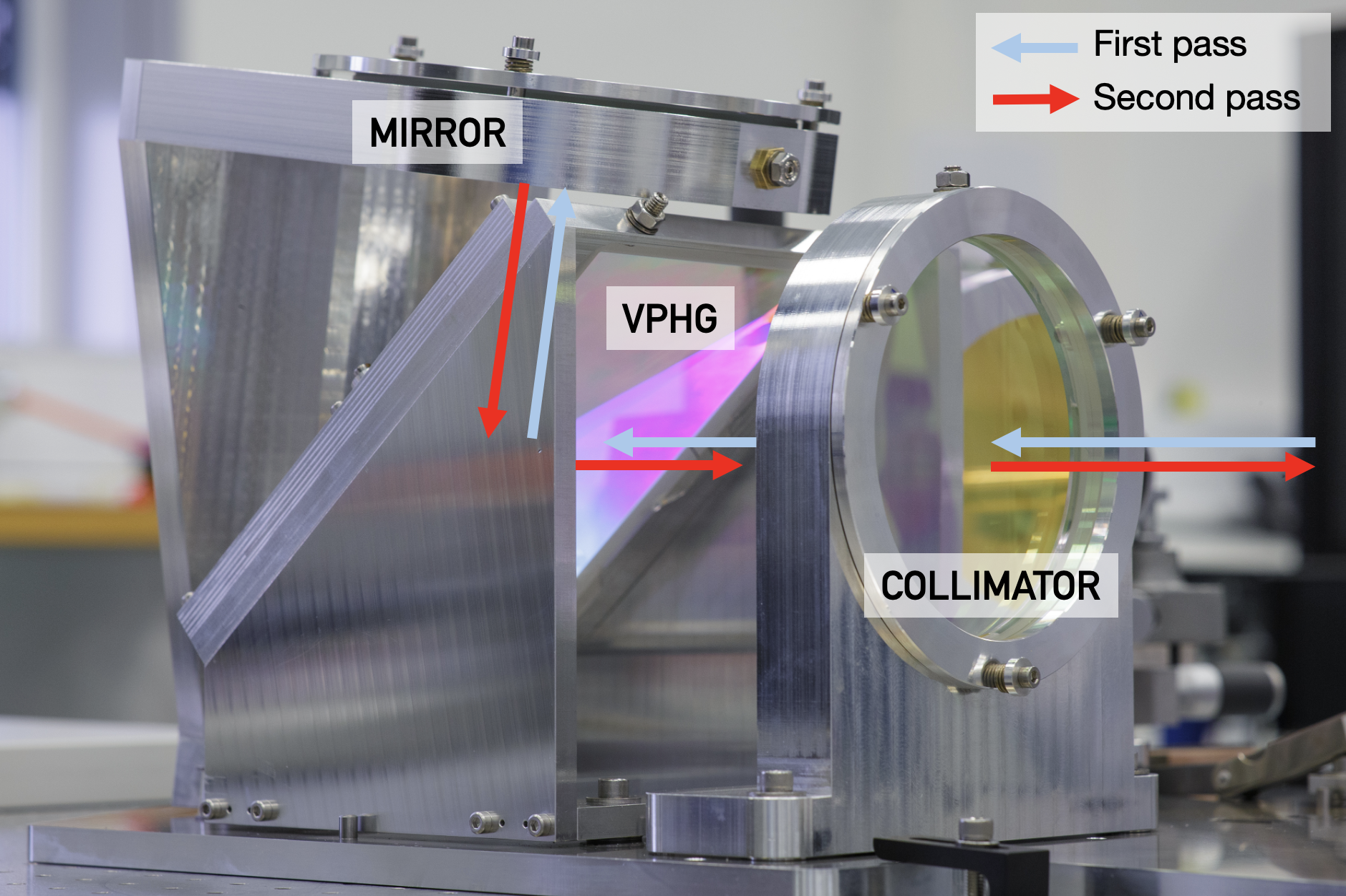}
\caption{In this picture we show the layout of the VPHg in the NIGHT instrument. The lens on the right-hand-side is the collimator. After light has passed through the collimator, it hits the grating, which is inclined at $49^\circ$. Above the grating rests a mirror which reflects light back onto the grating. After the second pass through the grating, light passes in reverse through the collimator, in almost the same direction as light originally came in. The entrance and exit beams are slightly offset from each other by a $1^\circ$ tilt to separate the exit and entrance pupils of the spectrograph. This ensures that the final spectrum can be sent to the detector, rather than being returned to the location where light is injected into the spectrograph.}
\label{fig:night_vph}
\end{figure}

\begin{figure}[!ht]
\centering
\includegraphics[width=0.65\textwidth]{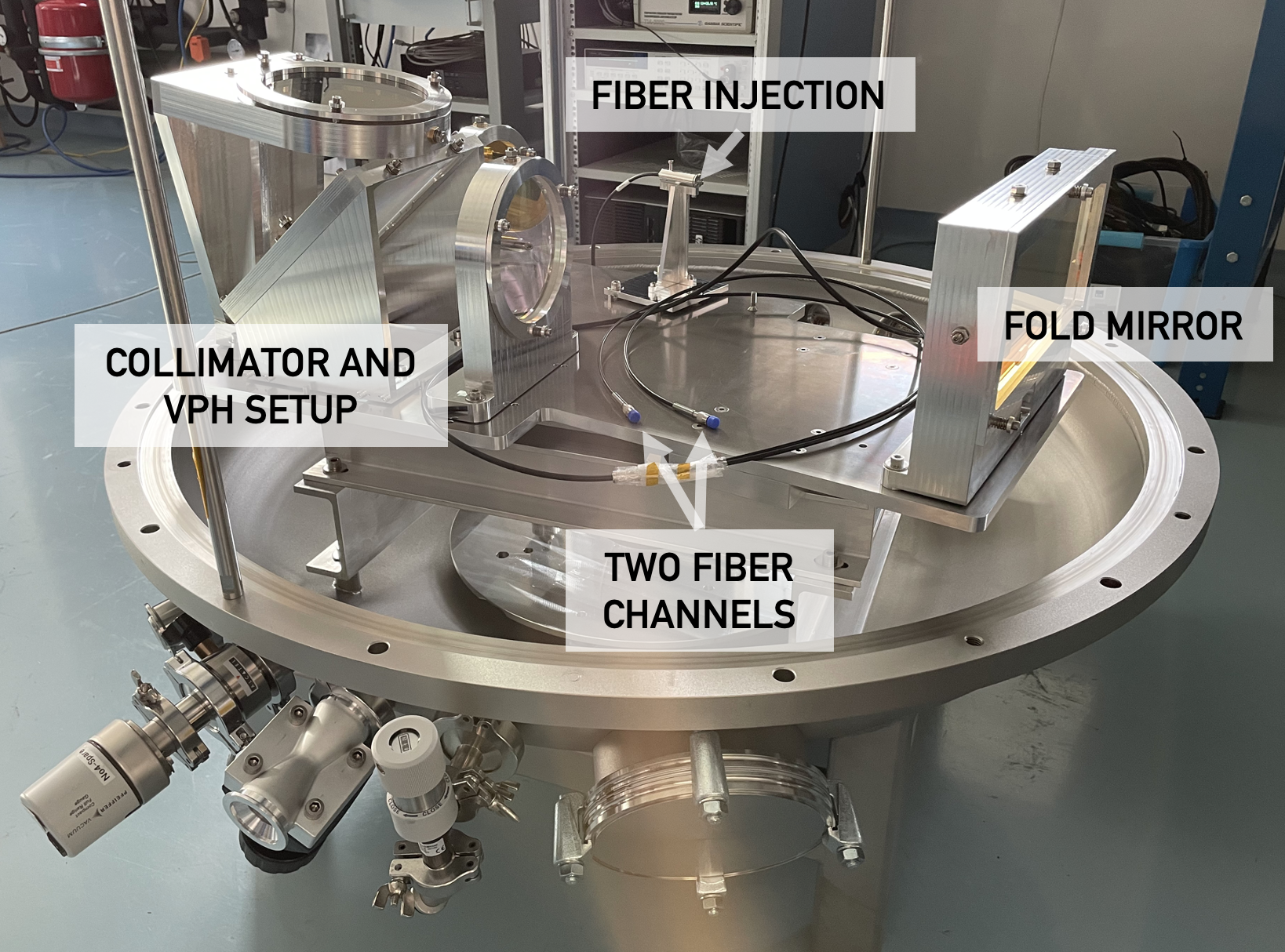}
\caption{This picture depicts the NIGHT spectrograph with its vacuum vessel bell-housing removed. The VPHg setup illustrated in Fig.~\ref{fig:night_vph} is visible on the left-hand side of the optical bench. At the rear, one can observe the pillar that houses the fiber injection system for both channels. The fold mirror positioned on the right-hand side operates in a triple-pass configuration and features a custom dielectric anti-reflective coating that reduces optical losses to negligible levels.}
\label{fig:night_spectro}
\end{figure}

Almost all astronomical high-resolution ($\mathrm{R}>50,000$) spectrographs have been designed with an echelle grating as the main disperser. The echelle, with its operation at high spectral order, enables simultaneous broad wavelength coverage and high resolving power. Both are essential for most exoplanet-related science—for example, RV measurements require numerous resolved spectral lines to reach high RV precision. For NIGHT, the focus on the helium triplet allowed us to relax one of these two requirements: a broad spectral coverage is not needed. This made us decide to move away from the echelle grating and look for a more efficient alternative. The answer was found in the Volume\,Phase\,Holographic grating (VPHg), a grating type that has many advantages over an echelle: better optical quality, higher diffraction efficiency, and improved handling. A custom VPHg was designed and made for NIGHT with an efficiency of $\sim 90\%$. To reach the spectral resolution requirement of NIGHT, the VPHg had to be placed in a double-pass configuration (resulting in $\sim 81\%$ diffraction efficiency, almost double the peak efficiency of a typical cross-dispersed echelle set-up), allowing us to reach the highest resolving power ever achieved with a single VPHg \cite{FarretJentink2025} (see Fig.~\ref{fig:night_vph}). The full spectrograph's optics are shown in Fig.~\ref{fig:night_spectro}.

\subparagraph{The Front End}

\begin{figure}[!ht]
\sidecaption
\includegraphics[width=0.5\textwidth]{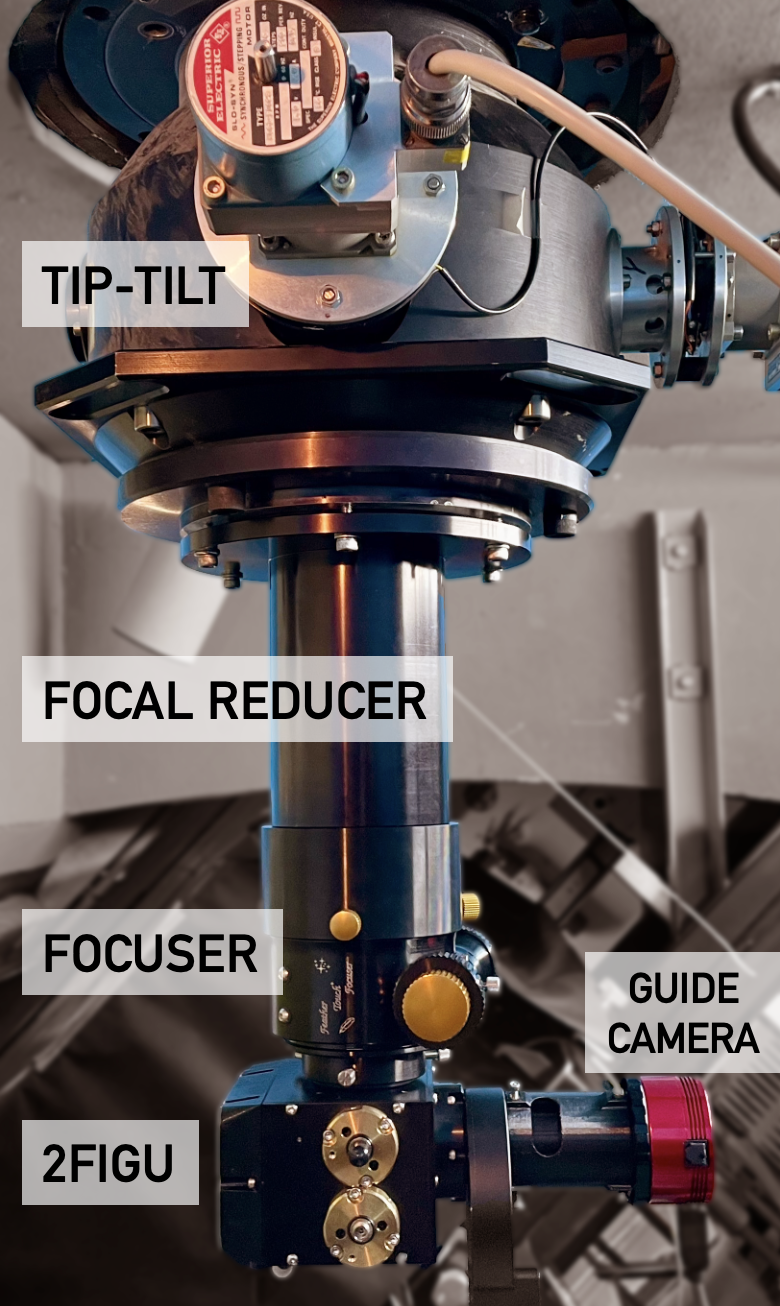}
\caption{The NIGHT front-end guide and injection unit installed on the 152cm telescope of Observatoire Haute Provence. The entire assembly is mounted at the Coudé focus of the telescope. Starlight enters from the telescope at the top of the image. It initially passes through a tip-tilt plate controlled by two stepper motors--a component that was pre-installed at the telescope before our arrival, but which we interface with for active guiding. Following the tip-tilt plate, the light traverses our custom 3-element focal reducer, then an off-the-shelf (OTS) but modified focuser, and finally enters the 2FIGU. The guide camera employs an OTS \textsc{ZWO} infrared-sensitive detector paired with an infrared filter, which effectively eliminates the need for an atmospheric dispersion corrector.}
\label{fig:NIGHT_frontend_OHP152}
\end{figure}

To ensure stability and precision of NIGHT, we implemented a fiber-fed design. Optical fibers effectively homogenize the light distribution at the spectrograph's entrance slit--accomplished through non-circular fibers and a double scrambler that interchanges the near- and far-field. The injection of starlight from a telescope into optical fibers necessitates a fiber injection system, a feature common among high-resolution spectrographs.
Typically, such systems combine an optical fiber with a pierced mirror and a guide camera focused on this mirror. The fiber is positioned behind the mirror's pierced aperture, while the guide camera monitors whether starlight properly falls on this aperture. When starlight deviates from the hole, the offset is automatically converted to telescope coordinates and transmitted to a correction system that adjusts telescope pointing--a process known as active guiding.
Following this same approach, we designed and constructed a custom fiber-injection and guide unit for NIGHT, designated as the 2FIGU (\textit{2-Fiber Guide Unit}), reflecting its dual optical fiber configuration. The 2FIGU has two fibers, hereafter defined as channels A and B, to be able to simultaneously observe sky emission besides starlight. The helium triplet is known to be quite spectrally close to an OH telluric emission line. As such, we observe the sky in a separate channel to aid in data reduction during post-processing of the spectra.

In addition to two output fibers, the 2FIGU also has the option to inject calibration light into channels A and B. The 2FIGU has a third fiber connector port on its outer housing to couple calibration light. An internal motorized slider mechanism can position the image of this calibration fiber over either hole of the pierced mirror, sending calibration light into the channels leading to the NIGHT spectrograph (also see Fig.~\ref{fig:systems_sketch_NIGHT} and~\ref{fig:NIGHT_frontend_OHP152}). 

\section{Implementation of NIGHT at the 152cm Telescope}

NIGHT will be installed at the OHP152 for a period of less than a year, from early summer 2025 to March 2026. After this period, the telescope will be decomissioned. Given its last year of operation and its age, there is little time pressure on the OHP152, making it ideal for validation of NIGHT. 

\begin{figure}[!ht]
\centering
\includegraphics[width=0.9\textwidth]{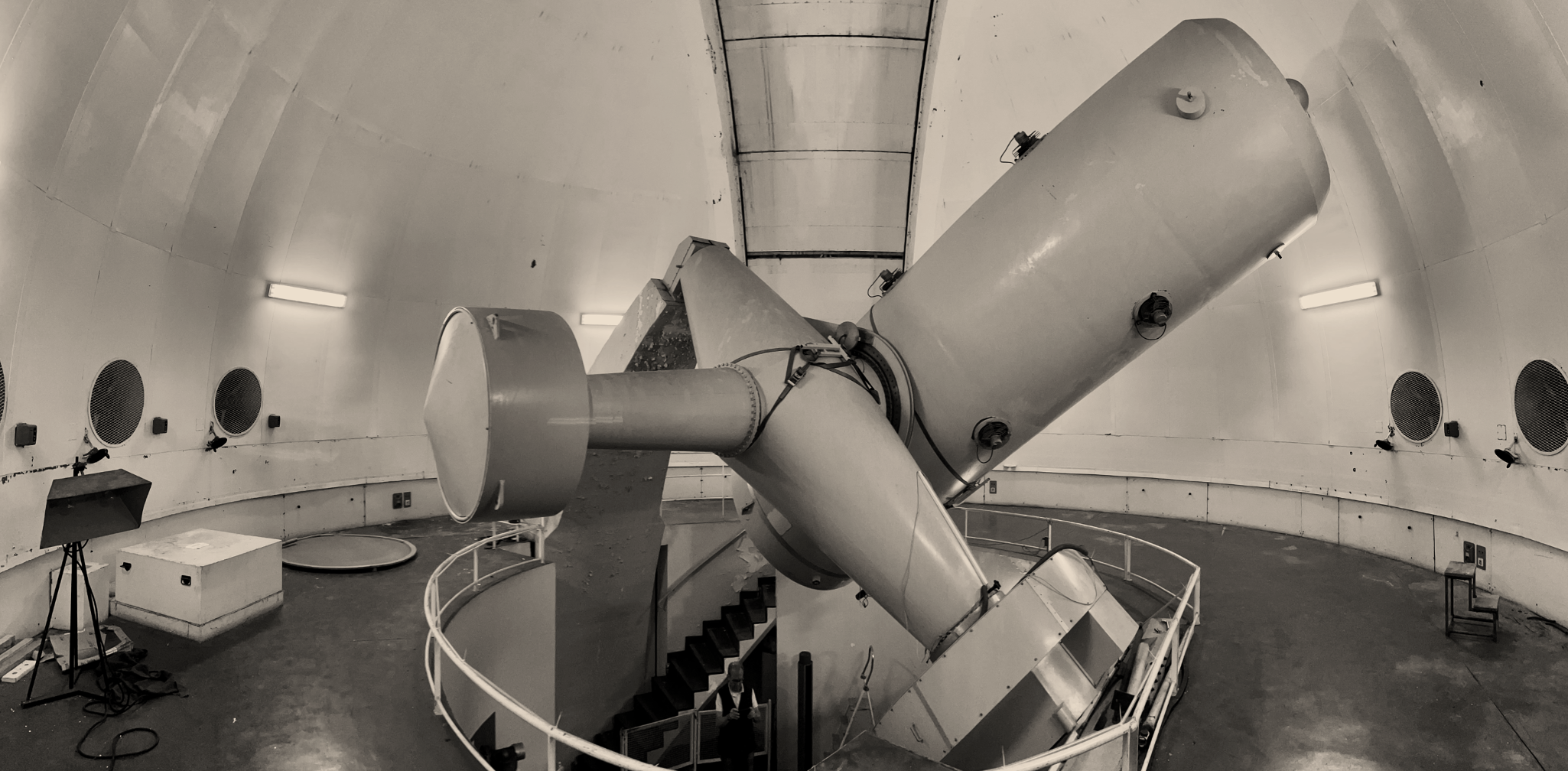}
\caption{The 152cm telescope at Observatoire Haute Provence was constructed in the late 1960s. At the end of February 2026, the telescope will be decommissioned, after almost 60 years of service. }
\label{fig:theOHP152}
\end{figure}

\begin{figure}[!ht]
\centering
\includegraphics[width=0.99\textwidth]{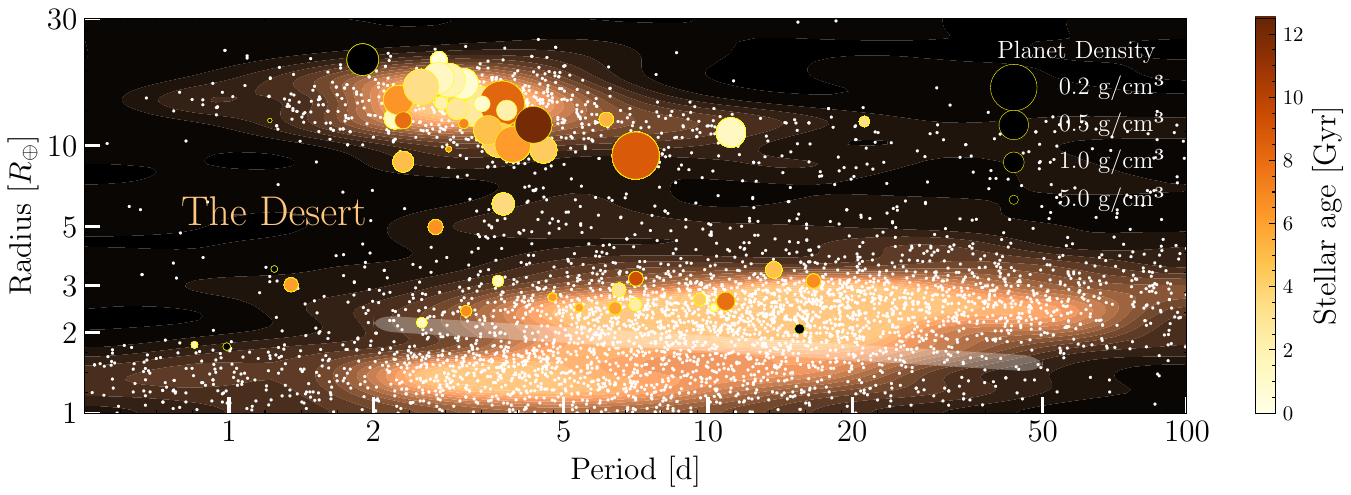}
\caption{The planets targeted by NIGHT during its inaugural year at the OHP 152cm telescope are represented as yellow-bordered circular markers as a function of period and radius. The marker colors indicate the age of each planetary system, though it should be noted that stellar ages typically carry substantial uncertainties. Black markers denote systems for which no age determination is available. Marker size corresponds to the average planetary density, with larger markers representing lower-density worlds. The small white circular markers are all known transiting exoplanets. The white semi-transparent region illustrates the location of the Radius Valley according to the definition established by \cite{ho2023}.}
\label{fig:NIGHT_targets}
\end{figure}

Given the telescope's age and limited automation, an on-site observer is required. This necessity imposes significant constraints on observation periods. As our team lacks the human resources to maintain an observer every night, we established a series of dedicated observing runs. Through careful analysis--considering minimum target magnitude (J$<$12) required for guiding and excluding planets where atmospheric escape appears highly implausible ($\mathrm{T_{eq}<300 K }$)--we evaluated transit visibility for all remaining known exoplanets during the period from April 2025 to May 2026 at high-enough signal-to-noise (S/N). This process yielded approximately 50 candidate planets, some with multiple transits (see Fig.~\ref{fig:NIGHT_targets} for the period-radius plot). We scheduled 35 of these planets for observation across six dedicated runs.

The final step in the scheduling process was a trade-off of overlapping transits, where transits for different targets happen on the same night. Here, we tried to keep as many targets as possible. As an example, we would favour discarding a planet if it was observed during another day in our pre-defined runs. However, in the cases where we had to choose between two planets, preference for observation was given to puffier planets, closer to the Neptunian Desert and Radius Valley. 

Although this selection method is not without bias--it already starts with the inevitable magnitude and S/N limitations and inherent bias to planets on shorter orbital periods, as these transit more frequently--we believe that it is as close as we can get to an unbiased, homogenous sample. 

\section{Computing transmission spectra from high-resolution spectroscopic data}

\subsection{Probing atmospheres through transmission spectra}

Transiting exoplanets cause a dimming of the star’s light, typically ranging from $\sim$1\% for gas giants down to $\sim$80\,ppm for Earth-sized planets. These values, however, relate to the bulk of the planet, optically thick to the stellar light across broad spectral ranges. A much larger fraction of stellar light can be absorbed at specific wavelengths by species with strong absorption cross-sections, present at high altitudes in the optically thin layers of the planetary atmosphere. This filtering of the stellar light by the atmospheric limb of a transiting planet has proven particularly effective for probing the composition, dynamics, and physical structure of the different atmospheric layers (\cite{seager2000,Seidel2025}). At first order, the amplitude of the absorption signal from the optically thin layers scales with their temperature, mean atmospheric molar mass, and planet bulk density. This scaling favours the hydrogen-rich atmospheres of hot Jupiters, which were the first class of planets from which escape was detected.

Soon after the first observation of an exoplanet atmosphere\cite{charbonneau2000}, the UV spectroscopic capabilities of the Hubble Space Telescope allowed detecting the neutral hydrogen exosphere of a hot Jupiter through its absorption of the Lyman-$\alpha$ transition\cite{Vidal2003}. Although UV observations led to further detections over the following decade, in particular for less irradiated Jupiters (\cite{Ehrenreich2012}), smaller Neptune-size planets (eg, \cite{Ehrenreich2015,Bourrier2018b}), and through the absorption from metal species (eg, \cite{Sing2019}), the difficulties associated with Lyman-$\alpha$ observations prevented searching for atmospheric escape in a large sample of planets (\cite{Owen2023}).   

In parallel to space-based observations, ground-based spectrographs - originally designed to detect exoplanets through radial velocity monitoring - progressively offered an increasingly detailed window into exoplanetary atmospheres. High spectral resolution allows distinguishing between the stellar, planetary, and telluric lines. The first detections probed the lower atmospheric layers, by tracking molecular bands from water and carbon monoxide at infrared wavelengths (e.g. \cite{Snellen2010,Brogi2012}). This method was then progressively extended to the optical domain, resolving atomic absorption lines at higher altitudes in hot gas giant atmospheres (\cite{Wyttenbach2015}), and giving access to the measurements of temperature gradients and atmospheric dynamics at the transition between the stratosphere and thermosphere (\cite{Wyttenbach2017,Seidel2019,Seidel2020a}). Iconic results based on the Na I doublet include the detection of significant blueshifts in hot Jupiters (e.g. \cite{Wyttenbach2015,Langeveld2022}) and ultra-hot Jupiters (e.g., \cite{Seidel2025}), as well as a broadening of the lines (e.g., \cite{Seidel2019}). These features were attributed to high-velocity flows from the hot star-facing side to the cooler space-facing side of the planet (e.g., \cite{Seidel2023}), as expected from the region at the base of the thermosphere where hydrodynamic expansion is given birth (e.g., \cite{Vidal2003,lammer2003,koskinen2022,Salz2016}). These results illustrate how high-resolution ground-based observations at optical wavelengths started bridging the gap between the lower atmospheric layers and escaping exosphere. 
In complement to optical observations, ground-based spectroscopy in the near-infrared recently opened a new window into the transition between the thermosphere and exosphere. The first detection of metastable helium \cite{seager2000,oklopcic2018,spake2018} in an exoplanet atmosphere showed its potential to determine mass-loss rates in a much larger sample of planets than previously accessible via UV spectroscopy. 

\subsection{The challenges of transmission spectra}

What is measured during an in-transit observation is the spectrum integrated over the full stellar disk, which misses the contribution fully hidden by the opaque planetary layers, and contains the light filtered by the limb of the planetary atmosphere. The rest - the light emitted by the regions that are not occulted by the planet - can be removed through subtraction with the stellar spectrum measured outside of the transit. To then isolate the transmission spectrum of the planetary atmosphere, we must remove by division the spectrum emitted by the annular region of stellar surface filtered by the planetary limb during a given observation. Finally, transmission spectra extracted during individual in-transit exposures generally need to be cumulated to gain a sufficient S/N for analysis, which requires aligning them in the common rest frame of the planetary atmosphere.

In practice, we do not have access to the spectrum occulted locally by the planet, since stellar surfaces cannot be spatially resolved with current instrumentation. An accurate proxy for the local stellar spectrum occulted in each exposure must thus be estimated, which requires determining the shape of the stellar line profiles and their radial velocity position in the star rest frame. Originally, these limitations were neglected and mitigated when probing exospheres, using the disk-integrated stellar spectrum as a proxy for the local one. This is because the large spatial extension and fast dynamics of the escaping gas yield broad spectral features (e.g., \cite{DosSantos2019,Bourrier2020_MOVESIII}), which are further blurred by the use of long exposure durations and the medium-spectral-resolution STIS and \textit{Cosmic Originis Spectrograph} (COS) spectrographs on HST (e.g., \cite{Carteret2024})

The advent of high-resolution spectrographs, in particular HARPS\cite{pepe2002} and then ESPRESSO\cite{pepe2010}, greatly improved the precision of transmission spectra but also proved extremely sensitive to the planet-occulted line distortions (POLDs, \cite{Dethier2023}) introduced by dividing the true planet-occulted stellar lines with the disk-integrated ones. Only when a star rotates slowly (limiting rotational broadening of the disk-integrated lines and the radial velocity offset of the occulted lines) and shows no strong center-to-limb variations of the local lines can the POLDs be neglected. In cases where the planetary line(s) probed are also present in the stellar spectrum, POLDs contamine the narrow absorption lines from the planetary atmosphere (e.g. \cite{Casasayas2020,CasasayasBarris2021,Yan2023}). The helium triplet lines are present in most stellar types (e.g., \cite{smith2016, sanz-forcada2008, lambert1987, obrien1986}), and thus require correction. Magnetic activity from the star can further introduce inhomogeneities in the spectral emission of the stellar surface, making it even more critical to determine the exact local spectrum occulted by the planet (\cite{Boldt2020}).

While techniques have been developed to correct for POLDs a posteriori in transmission spectra, such corrections are mathematically incorrect and introduce biases in the derived planetary spectra (\cite{Dethier2024}). One needs to either determine an accurate proxy for the occulted stellar lines to isolate the pure transmission spectrum of the planetary atmosphere, or to simulate realistic transmission spectra containing both POLDs and planetary atmospheric signatures. 

\subsection{The interpretation of transmission spectra}

Both approaches - isolating planetary transmission spectra or modelling POLD-contaminated transmission spectra -  require determining precisely the RV shifts of the occulted stellar lines, which can be determined through analysis of the Rossiter-McLaughlin (RM) effect (e.g., \cite{Triaud2009}). This technique relies on the occultation of the local stellar regions by the opaque layers of the planet, using spectral lines that are not affected by absorption from the optically thin planetary atmosphere. These lines can be cross-correlated into a ``white-light'' average line, or CCF, that can be analyzed more directly. Traditionally, the analysis of the RM effect consisted of studying the distortion of the disk-integrated CCFs induced by the occulted local CCF (e.g., \cite{Queloz2001}). More recent techniques rely on the extraction of the occulted CCFs and the analysis of their centroid (\cite{Cegla2016}) or full spectral profile (\cite{Bourrier2021}). All RM techniques allow constraining the stellar rotational velocity and spin-orbit angle of the planetary orbit which, combined with the planet impact parameter, yield the RV of the local stellar surface along the transit chord. 

Both approaches further require using a better proxy for the occulted stellar lines than the disk-integrated ones. A higher accuracy for these proxies is required to isolate planetary transmission spectra. As explained in the next section, using an approximate proxy for the occulted stellar line is less critical when they are used to compute numerical stellar spectra, and then transmission spectra, comparable to observations.  

In any case, the computation of transmission spectra requires a thorough and robust processing of the measured stellar spectra. As the number of systems observed with different instruments kept growing, it indeed became crucial to develop a standard in analysing spectral time-series of exoplanet transits for the benefit of reproducibility. In that spirit, the open-source\footnote{\url{https://gitlab.unige.ch/spice\_dune/antaress}} ANTARESS workflow (\cite{Bourrier2024}) developed within the frame of \textit{PlanetS} provides a set of methods to process high-resolution spectroscopy datasets. In the present version, echelle spectra from multiple instruments and epochs can be corrected for relevant environmental and instrumental effects, processed homogeneously, and analysed independently or jointly. Planet-occulted stellar spectra extracted along the transit chord enable a spectral and spatial mapping of the photosphere (which can be used to analyse the RM effect) and provide direct comparison with theoretical stellar models. Cleaned, disc-integrated spectra can then be used for global stellar characterisation, and to compute state-of-the art transmission spectra that can be used for forward modelling of POLDs and atmospheric signatures.

\section{Interpreting high-resolution transmission spectra with atmospheric escape models}

As highlighted in the previous section, the comparison between models and observations can be challenging. Indeed, transmission spectra are inherently linked to the occulted local stellar surface. When not accounted for, stellar surface inhomogeneities can mimic an additional absorption signature that is not related to the planetary atmosphere. This will ultimately create a bias in the retrieved properties of the atmosphere, highlighted by recent works \cite{Dethier2023,Carteret2024,Carteret2025} including members of \textit{PlanetS}. As such, a rigorous treatment of both the star and planet, and an adequate observational strategy are necessary to properly assess the structure and dynamics of the upper atmosphere. 

\subparagraph{Upper atmosphere modelling}

The upper layers of hot giant planets are primarily composed of hydrogen and helium (typical abundances used in the literature vary between 90-10\% and 99-1\% depending on the planet). Due to their close proximity with the star, the hydrogen and helium in the atmosphere are highly irradiated and heated. As a result the thermosphere inflates, atoms accelerate and can be brought beyond the Roche lobe, causing atmospheric escape \cite{Yelle2004,Tian2005,Owen2012}. The thermosphere is dense enough to be accurately described using hydrodynamic models. Indeed, the Knudsen number, defined as the ratio between the mean free path of particles and the scale height, remains well below 1. In contrast, the exosphere, which extends farther from the planet, is collisionless and depleted through photo-ionization and interactions with the wind of the host star. The transition between these two regions remains poorly understood, and often assumed (by default) to occur at the planetary Roche lobe. Ever since the first detection of atmospheric escape from an exoplanet there has been an ongoing debate over whether the exosphere can be well described within the hydrodynamic framework, especially for outflows extending far from the planet.

Given the high abundance of neutral hydrogen in the upper atmospheres of close-in giants, and strong oscillator strength of the Lyman-$\alpha$ transition, the resulting absorption signal was expected by scientists to be the strongest among atmospheric escape tracers. The absorption signature of neutral hydrogen's Lyman-$\alpha$ line \cite{Vidal2003} was observed very early in exoplanetary research, as described in the previous sections. This observable allows for indirect estimation of total atmospheric mass-loss rates and has paved the way for the field of atmospheric escape studies. Lyman-$\alpha$ transit signatures typically extend beyond the nominal transit duration, tracing the presence of hydrogen atoms at large distances from the planet. In the most extreme cases, the stellar XUV flux is insufficient to fully ionize most of the escaping hydrogen within a few hours, allowing the outflow to extend to tens of planetary radii and cover a significant fraction of the orbit (see e.g. \cite{Lavie2017}). In the comet-like tail structures probed through Lyman-$\alpha$ observations, the hydrogen outflow density drops exponentially with distance from the planet, as expected from a regime with a large mean free path and sparse collisions. Naturally, researchers attempted to adapt codes originally developed for solar system physics, describing stellar wind as a fluid or simulating cometary tail with particle models, to interpret these observations. Consequently, there exists a dichotomy between two types of models developed after the first observations of exoplanetary escape. On one hand, hydrodynamical models assume that the outflow remains fully collisional on large scales \cite{Koskinen2013,Salz2016}. On the other hand, collision-less prescriptions assumed a particle exosphere located beyond the Roche lobe \cite{Volkov2011,Bourrier2013}. 

\begin{figure}[!ht]
\centering
\includegraphics[width=\textwidth]{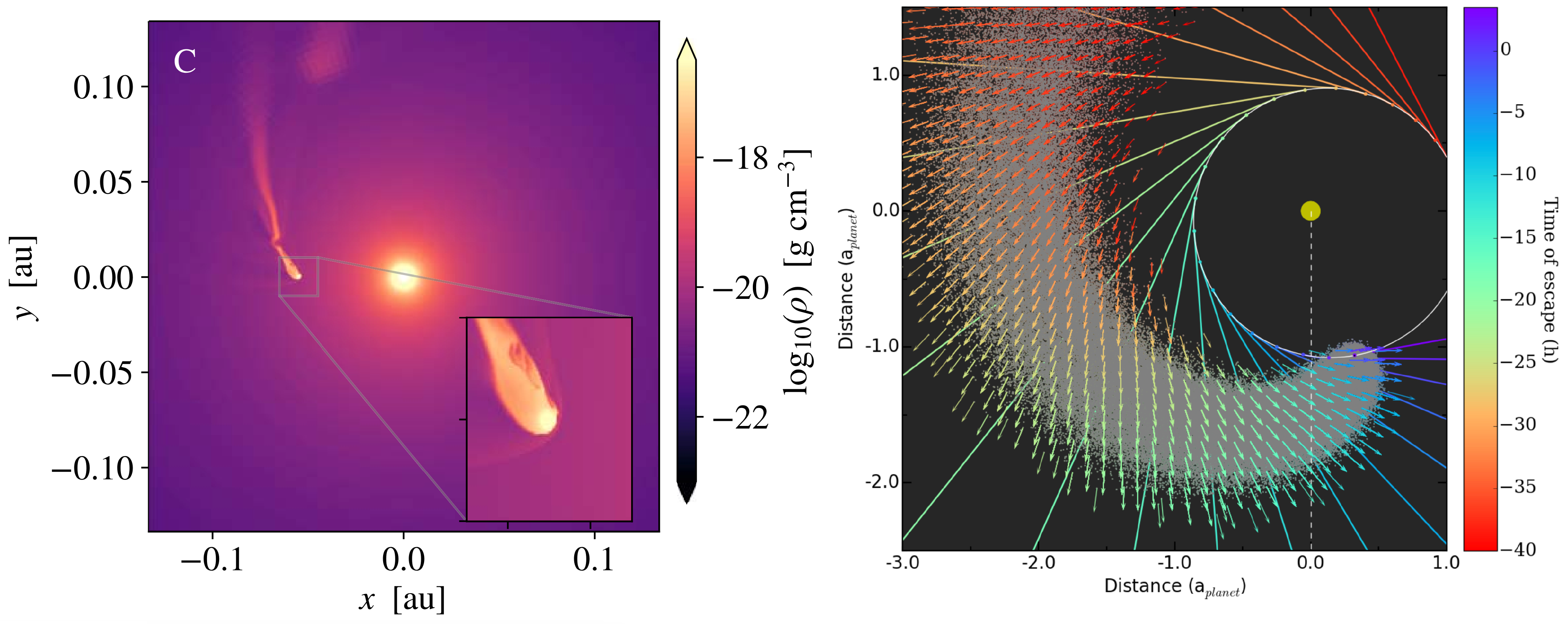}
\caption{Simulations of hydrodynamic (left, figure from \cite{MacLeod2022}) and non-collisional (right, figure from \cite{Bourrier2015}) outflows.}
\label{fig:hydrogen_tails}
\end{figure}

Since Lyman-$\alpha$ absorption primarily probes regions far away from the planet, the exact description of the atmosphere within the Roche lobe plays only a secondary role in interpreting the observed absorption features. As such, hydrogen absorption is controlled, at first order, by the density and velocity of the escaping flow at the Roche lobe, and its later interaction with the stellar environment. Due to the thermospheric expansion and the Roche potential, the escaping outflow typically forms trailing and leading tails with respect to the planet. However, the interaction with the stellar environment often prevents the formation of the leading tail. 
In the hydrodynamical prescription, the balance between the ram pressure of the stellar-winds and the planetary outflow strongly affects the structure of the tails. Collision-less models rather use stellar radiation pressure that acts very efficiently on isolated hydrogen atoms, pushing them away from the star at velocities of $\sim100$\,km\,$\cdot$\,s$^{-1}$ \cite{Bourrier2015,Beth2016}. The typical outflow structure obtained with both prescriptions is shown in Fig.~\ref{fig:hydrogen_tails}.   

As mentioned in previous sections, metastable helium has more recently emerged as a powerful tracer of atmospheric escape and does not provide a direct measurement of atmospheric mass-loss rate. In this context, the community has extensively developed thermospheric models including typical chemical reactions occurring within the upper atmosphere. The simplest of these approaches is the spherically symmetric isothermal Parker wind \cite{oklopcic2018,Lampon2020,DosSantos2022}. Building upon this first-order approach, the community started to integrate more physical processes in thermospheric models. \cite{Caldiroli2021} developed a fully hydrodynamical model, \cite{Allan2024,Schulik2024} included a more sophisticated metastable helium chemical network, \cite{linssen2024} extended the chemistry to a broader range of species, and \cite{Gillet2023} explored the impact of new processes such as secondary recombination, among others. Parker wind models treat the mass-loss rate as a free parameter, offering a large flexibility in reproducing the observational data. However, this comes at the cost of the lack of physical processes, limiting the detailed characterization of the mass-loss rate or mechanisms driving the atmospheric escape. In contrast, hydrodynamical thermospheric models are predictive by construction; the mass-loss rate is an output and depends on the specific physical assumptions implemented. This diversity of models available within the community highlights the difficulty in comparing mass-loss estimates across different studies, even with 1D models.

In parallel to the development of 1D thermospheric models, 3D fully hydrodynamical models have emerged to describe the large-scale structure of atmospheric outflows, encompassing metastable helium (\cite{Wang2021,MacLeod2022}). These simulations have shown that metastable helium can survive and escape on distances of several planetary radii in a hydrodynamical regime. Despite relying on simplified chemistry, these models include mechanisms that drive the shape of the outflow on large scales. It informs us on the interactions between the planetary outflow and stellar environment (stellar winds and tides), or its dependence on the thermospheric conditions (temperature gradient and sound speed for instance) \cite{MacLeod2022,macleod2024streams,nail2024cold}. The increased complexity introduced by these processes can make comparison with observations more difficult, and some specific features observed in transmission spectra remain unexplained \cite{Zhang2023,Gully_Santiago2024,Czesla2024}.

\subparagraph{Standard modelling}
This section highlights the standardized procedure for analysing atmospheric escape developed within the framework of \textit{PlanetS} that we propose to the community. 

As mentioned previously, local stellar inhomogeneities can affect transmission spectra. Using the Evaporating Exoplanets (EvE) code \cite{Bourrier2013,Bourrier2015}, we simultaneous model the planetary absorption and stellar contamination; fitting directly the time series of observed transmission spectra, rather than a single, time-averaged spectrum, naturally accounting for stellar inhomogeneities across the transit chord. EvE simulates planetary transit in 3D, accounting for the geometrical architecture of the system. The code produces time series of disk-integrated flux spectra at high temporal and spectral resolution. At each time step, the disk-integrated spectrum is computed from the local stellar spectra occulted by the opaque body, the thermosphere, and the exosphere together. In EvE, the thermosphere is treated as a fluid regime with a static grid, while the exosphere is collisionless, with its dynamics self-consistently updated at each time step. To construct the thermospheric grid, EvE interpolates density, velocity, and temperature profiles for each species onto an internal grid, providing flexibility in the choice of the atmospheric model. The properties of the particle exosphere, for example the escape rate, are computed directly from the thermospheric profiles at the exobase. The resulting time series of synthetic spectra can then be temporally resampled and convolved with an instrumental response to match an observed transit. 

In order to account for spatial variations of the stellar spectrum across the stellar disk, EvE uses a resolved stellar grid. Local stellar spectra are simulated using \texttt{Turbospectrum} \cite{plez1998,plez2012} in combination with a set of line lists, typically \cite{VALD,heiter2021,magg2022}. Additionally, the host star and orbital architecture of the system must be well-characterized, and prior RM, RVs and transits analyses are required to constrain the stellar rotation and the planetary orbit. We generate stellar grids with varying abundances of species strongly contributing to the stellar spectrum around the studied line, and fit them to the observed disk-integrated spectrum. The wavelength range of this fit should be chosen based on the observed velocity shift of the studied atmospheric escape tracer (due to its dynamics) across the full observational window. In the specific case of metastable helium, which is formed in the stellar chromosphere \cite{Andretta1997}, the triplet is not included in photospheric models. For this reason, we implemented an analytical model of the stellar metastable helium that approximates its absorption as Gaussians, parametrized by a column density and temperature\cite{dethi2023}. This model is then directly multiplied with the local stellar spectra generated with \texttt{Turbospectrum}. Stellar center-to-limb variations of the metastable lines are not accounted for in this procedure. However, this is a second order effect compared to limb darkening or the RM effect. 

In this framework, the XUV flux is also essential and must be modelled as accurately as possible. Indeed, metastable helium chemistry, both in the thermosphere and exosphere, is strongly driven by the XUV flux \cite{oklopcic2019}, and especially by the $5\text{--}504$~\AA range \cite{sanz-forcada2008}. However, the XUV is poorly constrained by observations and the community must often rely on scaling laws that provide rough broadband estimates \cite{linsky_intrinsic_2014}. These relations are also limited in spectral range, usually up to a ${\sim}1200$\,\AA. While this does not represent a major issue for thermosphere modelling, it has a serious impact on the photoionization rate of the exosphere as the cross-section is large at wavelengths ${>}1200$~\AA, overestimating the exospheric contribution. As a result, coronal models like \cite{Sanz_forcada_2025} are mandatory to properly study atmospheric escape. 

\subparagraph{Typical application}
We illustrate the proposed standard modelling with a concrete application from the first NIRPS atmospheric escape results on WASP-69\,b \cite{allart2025_1}.

\begin{figure}[!ht]
\centering
\includegraphics[width=\textwidth]{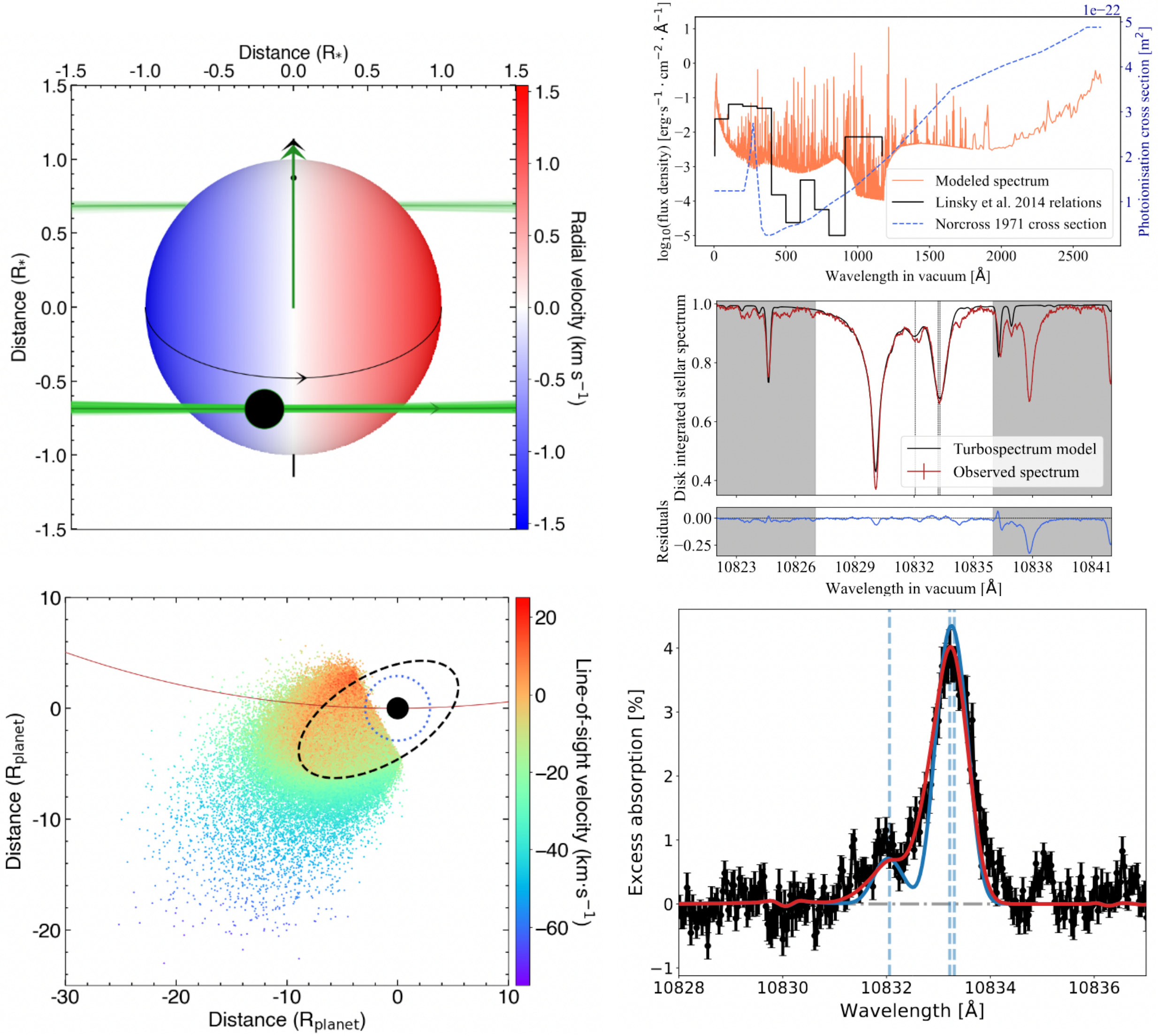}
\caption{All figures are taken from \cite{allart2025_1}, an example of the proposed standard modelling on WASP-69\,b. \textit{Top left: }Orbital architecture of WASP-69\,b. \textit{Bottom left: }Derived shape of the EvE thermosphere, together with the exospheric contribution. \textit{Top right: }Reconstructed stellar XUV and disk-integrated spectrum. \textit{Bottom right: }Observed and modelled average absorption spectrum.  }
\label{fig:wasp69}
\end{figure}

We observed 3 transits of WASP-69\,b with NIRPS and HARPS. The first step was to conduct the RM analysis over the entire dataset using the \texttt{ANTARESS} pipeline. It constrained the orbital architecture of the system as illustrated in Fig.~\ref{fig:wasp69}, specifically, we derived the projected stellar rotational velocity $v_\mathrm{eq} = 1.54 \pm 0.05$\,km$\cdot$s$^{-1}$ and projected spin-orbit angle $\lambda = 0.05 \pm 1.1$\,deg. Using the computed value for the rotational velocity, we then modelled the stellar spectrum, both in the XUV wavelengths and around the triplet, as described previously in this section. The stellar grid is matching reasonably well the observed disk-integrated spectrum, and allows for a description of local stellar inhomogeneities. The XUV spectrum was also constrained by previous observations, making it more representative of the true XUV received by the planet than scaling relations. The architecture of the system and stellar flux are then injected into EvE to simulate a transit, accounting for the stellar contamination, planetary opaque body, thermosphere, and exosphere. We performed a fit over the 3 transits time-series and derived a possible geometrical configuration of the planetary outflow taking the shape of an ellipsoid extending up to 10 times the planetary radius see Fig.~\ref{fig:wasp69}. The absorption spectrum was also well reproduced, constraining the mass-loss rate to $\dot{\mathrm{M}}\sim 2.25\cdot10^{11}$\,g/s. The H/He ratio was set as a free parameter in this study, favoring values between 0.9 and 0.8. This study highlights the workflow that we propose to the community to become a standardized approach to all atmospheric escape analyses.

\section{The future of atmospheric escape}

\subsection{NIGHT's dawn}

NIGHT is poised to significantly advance our understanding of helium atmospheres across diverse planetary types. We project a potential 50\% increase in metastable helium detections within the first year of operations. Systematic data retrieval and interpretation through ANTARESS and EvE will yield a quantitatively comparable dataset, illuminating which planetary classes exhibit atmospheric escape and at what rates. This comprehensive survey could provide robust observational constraints on the formation and evolution mechanisms related to atmospheric escape underlying the Neptunian Desert and Radius Valley. The project's approach to target selection and data processing will enhance our ability to qualitatively compare and combine multiple transit observations and optimize out-of-transit baseline selection. Furthermore, NIGHT will identify promising candidates among faint targets or with extended atmospheric tails for space-based follow-up with JWST. After its first operational period at the OHP 152cm telescope, we plan to move NIGHT to a larger telescope for increased sensitivity. While helium has emerged as a powerful tracer for probing exoplanetary upper atmospheres during the \textit{NCCR PlanetS}, its interpretation depends on various theoretical frameworks and models. Space-based UV observations would provide valuable complementary constraints to these models.

\subsection{A call for new UV missions}
Helium serves as a versatile atmospheric tracer detectable via ground-based observations, for example with NIGHT. However, interpreting these observations in the context of atmospheric escape presents significant challenges, requiring multiple models with poorly constrained parameters such as hydrogen-helium fractionation and stellar host XUV flux. To enhance model constraints, XUV to NUV spectroscopy offers critical information, yet these wavelengths are completely absorbed by Earth's atmosphere, necessitating space-based telescopes.

Several active space missions currently operate across portions of this spectral range. For X-ray observations, XMM Newton provides spectroscopic capabilities in the 0.5--3.5 nm regime\cite{Jansen2001}, while the Chandra X-Ray Observatory covers wavelengths from 0.12--12 nm\cite{weisskopf2000}. At longer wavelengths, the HST offers both the COS instrument\cite{green2012}, sensitive down to 81.5 nm, and STIS\cite{woodgate1998}, detecting wavelengths to 115 nm.

A substantial gap exists between these observatories' wavelength ranges--a region crucial for constraining stellar activity yet remains poorly understood \cite{sanz-forcada2008, chadney2015}. Though historical missions such as ROSAT \cite{Truemper1982} have explored portions of this regime, and new missions like ESCAPE \cite{France2019} have been proposed, this spectral region remains inaccessible.

In the longer-wavelength UV domain, HST stands as the sole active telescope with sufficient sensitivity to characterize exoplanetary atmospheres. Despite its aging infrastructure, no immediate successor is planned. The upcoming NASA UVEX mission could provide access to these wavelength regimes, but exoplanetary atmospheric characterization is not one of its prime science objectives \cite{kulkarni2012}. The UV wavelength band accessible via HST contains promising metal absorption lines that would complement metastable helium observations, enabling probes of different atmospheric depths and facilitating retrieval of accurate temperature and velocity profiles in the upper atmosphere \cite{Linssen2023}. Newer, smaller missions such as the CUTE CubeSat telescope have shown great potential to continue providing access to the UV domain in the characterization of exoplanet atmospheres.\cite{Fleming2018} The newest initiative is an ESA mission proposal named WALTzER, with strong contributions from PlanetS researchers, which could potentially observe the skies from the 2030s.

\newpage

\bibliographystyle{unsrt}

\bibliography{refs}

\end{document}